\documentclass[aps,prc,groupedaddress,showpacs]{revtex4-1}
\usepackage{amsfonts}

\usepackage{graphicx}
\usepackage{color}
\usepackage{longtable}
\usepackage{amssymb}
\usepackage{supertabular}
\usepackage{dcolumn}

\renewcommand{\baselinestretch}{1.5}

\begin{document}

\preprint{APS/123-QED}

\title{$\gamma$-ray linear polarization measurements and $(g_{9/2})^{-3}$ neutron alignment in $^{91}$Ru}
\author{Y.~Zheng$^{1,2}$}
\email[]{zhengyong@impcas.ac.cn}
\author{G.~de France$^{1}$}
\author{E.~Cl\'{e}ment$^{1}$}
\author{A.~Dijon$^{1}$}
\author{B.~Cederwall$^{3}$}
\author{R.~Wadsworth$^{4}$}
\author{T.~B\"{a}ck$^{3}$}
\author{F.~Ghazi Moradi$^{3}$}
\author{G.~Jaworski$^{5,6}$}
\author{B.M.~Nyak\'{o}$^{7}$}
\author{J.~Nyberg$^{8}$}
\author{M.~Palacz$^{6}$}
\author{H.~Al-Azri$^{4}$}
\author{G.~de Angelis$^{9}$}
\author{A.~Atac$^{3,10}$}
\author{\"{O}.~Akta\c{s}$^{11}$}
\author{S.~Bhattacharyya$^{1}$}
\author{T.~Brock$^{4}$}
\author{P.J.~Davies$^{4}$}
\author{A.~Di Nitto$^{12,20}$}
\author{Zs.~Dombradi$^{7}$}
\author{A.~Gadea$^{13}$}
\author{J.~Gal$^{7}$}
\author{P.~Joshi$^{4}$}
\author{K.~Juhasz$^{14}$}
\author{R.~Julin$^{15}$}
\author{A.~Jungclaus$^{16}$}
\author{G.~Kalinka$^{7}$}
\author{J.~Kownacki$^{6}$}
\author{G.~La Rana$^{12}$}
\author{S.M.~Lenzi$^{17}$}
\author{J.~Molnar$^{7}$}
\author{R.~Moro$^{12}$}
\author{D.R.~Napoli$^{9}$}
\author{B.S.~Nara Singh$^{4}$}
\author{A.~Persson$^{3}$}
\author{F.~Recchia$^{17}$}
\author{M.~Sandzelius$^{3}$}
\author{J.-N.~Scheurer$^{18}$}
\author{G.~Sletten$^{19}$}
\author{D.~Sohler$^{7}$}
\author{P.-A.~S\"{o}derstr\"{o}m$^{8}$}
\author{M.J.~Taylor$^{4,21}$}
\author{J.~Timar$^{7}$}
\author{J.J.~Valiente-Dobon$^{9}$}
\author{E.~Vardaci$^{12}$}

\affiliation{
$^{1}$ Grand Acc\'{e}l\'{e}rateur National d'Ions Lourds (GANIL), CEA/DSM - CNRS/IN2P3,
Bd Henri Becquerel, BP 55027, F-14076 Caen Cedex 5, France. \\
$^{2}$ Institute of Modern Physics, Chinese Academy of Sciences, Lanzhou, China \\
$^{3}$ Department of Physics, Royal Institute of Technology, SE-1069 Stockholm, Sweden. \\
$^{4}$ Department of Physics, University of York, YO10 5DD York, UK \
$^{5}$ Faculty of Physics, Warsaw University of Technology, Koszykowa 75, 00-662, Warsaw, Poland. \\
$^{6}$ Heavy Ion Laboratory, University of Warsaw, ul. Pasteura 5a, 02-093, Warsaw, Poland. \\
$^{7}$ MTA ATOMKI, H-4001 Debrecen, Hungary. \\
$^{8}$ Department of Physics and Astronomy, Uppsala University, SE-75120 Uppsala, Sweden. \\
$^{9}$ Instituto Nazionale di Fisica Nucleare, Laboratori Nazionali di Legnaro, I-35020 Legnaro, Italy. \\
$^{10}$ Department of Physics, Ankara University, 06100 Tandogan Ankara, Turkey. \\
$^{11}$ Physics Department, Middle East Technical University, 06531 Ankara, Turkey. \\
$^{12}$ Dipartimento di Fisica, Universit\`{a} di Napoli and Istituto Nazionale di
Fisica Nucleare, I-80126 Napoli, Italy. \\
$^{13}$ IFIC, CSIC, University of Valencia, Valencia, Spain. \\
$^{14}$ Department of Information Technology, University of Debrecen, H-4010 Debrecen, Hungary. \\
$^{15}$ Department of Physics, University of Jyv\"{a}skyl\"{a}, FIN-4001 Jyv\"{a}skyl\"{a}, Finland. \\
$^{16}$ Instituto de Estructura de la Materia, CSIC, E-28006 Madrid, Spain. \\
$^{17}$ Diparimento di Fisica dell'Universit\`{a} di Padova and Instituto
Nazionale di Fisica Nucleare, Sezione di Padova, I-3512Padova, Italy. \\
$^{18}$ Universit\'{e} Bordeaux 1, CNRS/IN2P3, Centre d'Etudes Nucl\'{e}aires de Bordeaux Gradignan,
UMR 5797, Chemin du Solarium, BP120, 3317 Gradignan, France. \\
$^{19}$ The Niels Bohr Institute, University of Copenhagen, 2100 Copenhagen, Denmark. \\
$^{20}$ Institut f\"{u}r Kernchemie, Johannes G\"{u}tenberg-Universit\"{a}t Mainz,
Fritz Strassmann Weg 2, D-55128  Mainz, Germany. \\
$^{21}$ School of Physics and Astronomy, The University of Manchester, Manchester M13 9PL, UK. \\
}

\date{\today}

\begin{abstract}
Linear polarization measurements have been performed for
$\gamma$-rays in $^{91}$Ru produced with the $^{58}$Ni($^{36}$Ar,
$2p1n$$\gamma$)$^{91}$Ru reaction at a beam energy of 111 MeV. The
EXOGAM Ge clover array has been used to measure the
$\gamma$-$\gamma$ coincidences, $\gamma$-ray linear polarization and
$\gamma$-ray angular distributions. The polarization sensitivity of
the EXOGAM clover detectors acting as Compton polarimeters has been
determined in the energy range 0.3$-$1.3 MeV. Several transitions
have been observed for the first time. Measurements of linear
polarization and angular distribution have led to the firm
assignments of spin differences and parity of high-spin states in
$^{91}$Ru. More specifically, calculations using a semi-empirical
shell model were performed to understand the structures of the first
and second (21/2$^{+}$) and (17/2$^{+}$) levels. The results are in
good agreement with the experimental data, supporting the
interpretation of the non yrast (21/2$^{+}$) and (17/2$^{+}$) states
in terms of the $J_{\rm max}$ and $J_{\rm max}-2$ members of the
seniority-three $\nu(g_{9/2})^{-3}$ multiplet.
\end{abstract}


\maketitle

\section{Introduction}
The $Z>40$ $N = 47$ nuclei are three neutron holes below the $N=50$
closed shell. Their low-lying positive-parity level structure can be
interpreted in terms of the spherical shell model as an interplay
between proton-particle and neutron-hole excitations in the $g_{9/2}$
orbital. The possible excitations would then be those belonging to
the seniority-three configurations: $\nu(g_{9/2})^{-3}$, which can
generate spins up to 21/2$^{+}$ and
$\pi(g_{9/2})^{2}$$\nu(g_{9/2})^{-1}$, terminating at spin
25/2$^{+}$. The results of g-factor measurements for the lowest
8$^{+}$ state in the $N=48$ isotones $^{86}$Sr \cite{matthias75},
$^{88}$Zr and $^{90}$Mo \cite{hausser78} indicate that it
is essentially built from the alignment of a $g_{9/2}$
neutron pair with a small proton admixture which increases with $Z$.
Therefore, the neutron aligned
$\nu(g_{9/2})^{-3}_{J^{\pi}=21/2^{+}}$ state would be expected to be
yrast in the level structure of $N=47$ isotones.

The high-spin level structure of $^{89}$Mo ($Z=42$, $N=47$) has been
studied by M. Weiszflog et al. \cite{weiszflog93}. The Shell Model
interpretation performed with the code RITSSCHIL \cite{zwar85} and
within the $(p_{1/2}, g_{9/2})$ model space indicates that the
positive-parity states up to 25/2$^{+}$ mainly consist of the
proton aligned $\pi(g_{9/2})^{2}$$\nu(g_{9/2})^{-1}$ configuration.
A particularly interesting case is the one of the 21/2$^{+}$ state.
This state can be generated in the neutron fully aligned
$\nu(g_{9/2})^{-3}$ configuration but the calculations indicate that
this component is as small as $1\%$. This interpretation has been
confirmed by g-factor measurements of the 21/2$^{+}$ isomeric state
in $^{89}$Mo, proving the dominance of the $g_{9/2}$ proton
alignment \cite{weiszflog95}.

The trend observed in the $N=48$ isotones and the measurement in
$^{89}$Mo indicate an evolution from neutron to proton alignment, to
generate high-spin states in this mass region. In particular the
21/2$^{+}$ states in the $N=47$ isotone $^{91}$Ru might reveal a
complex structure. Understanding the microscopic structure of these
levels should therefore shed light on the competition between the
possible seniority schemes for the active $g_{9/2}$ protons and
neutrons.

Several groups have already studied the high-spin level structure of
$^{91}$Ru \cite{arnell93,heese94, dean04,gorska00}. Measurements
using $\beta$-decay, $\gamma$-$\gamma$ and $n$-$\gamma$ coincidences
as well as $\gamma$-ray anisotropy ratios have been performed and a
level scheme proposed. However, all the spin and parity assignments
were based on indirect evidences, systematics or Directional
Correlations of the $\gamma$-rays deexciting Oriented states (DCO
ratios) with fairly large uncertainties and had to be considered as
very tentative. The proper way to firmly assign a parity to an
excited state is to determine the electromagnetic character of the
transition deexciting this particular state. To do this, it is
necessary to measure its linear polarization. When combining the
polarization information with the angular distribution measurements,
the spins and parities of the excited states can be reliably
determined. In recent years, due to its high polarization
sensitivity and detection efficiency the Ge clover detector
\cite{Clover, Andgren07} has become a useful tool for the
measurement of linear polarization by using Compton scattering
between adjacent crystals.

In the present work we report on the results of linear polarization
measurements in $^{91}$Ru populated in the fusion-evaporation
reaction $^{58}$Ni( $^{36}$Ar, $2p1n$)$^{91}$Ru by using the EXOGAM
Ge clover detector array \cite{exogam}. As a result of this work,
non yrast (21/2$^{+}$) and (17/2$^{+}$) states have been observed
for the first time and added to the positive-parity structure of
$^{91}$Ru. A theoretical understanding of the structures of the
first and second (21/2$^{+}$) and (17/2$^{+}$) levels has been
obtained in terms of semi-empirical shell model calculations. In
addition, the polarization sensitivity of the EXOGAM clover
detectors, acting as Compton polarimeters, have been determined over
a wide range of $\gamma$-ray energies for the first time.

The paper is organized as follows: a description of the experiment
at GANIL and the data analysis with a special emphasis on the
polarization measurements and the first characterization of EXOGAM
as a Compton polarimeter will be presented in section II. In section
III, we present the results obtained for $^{91}$Ru while the Shell
Model calculations we performed to understand the microscopic nature
of the high-spin states in this nucleus will be discussed in section
IV.

\section{Experiment and data analysis}

Excited states in $^{91}$Ru have been investigated using the
fusion-evaporation reaction $^{58}$Ni($^{36}$Ar, $2p1n$) at a beam
energy of 111 MeV and with an average intensity of 10 pnA. The beam
was provided by the CIME cyclotron of GANIL, Caen, France. The
isotopically enriched (99.83\%) $^{58}$Ni targets used in the
reactions had an average thickness of 6.0 mg/cm$^{2}$, enough to stop
the recoiling nuclei. The $\gamma$-rays from the reaction products
were detected by the EXOGAM Ge clover detector array \cite{exogam},
consisting of 11 clover-type Ge detectors for this experiment, 7 at
an angle of 90$^\circ$ and 4 at 135$^\circ$ relative to the beam
direction. Neutrons evaporated from the compound nuclei were
detected using the Neutron Wall array \cite{skeppstedt99} composed
of 50 organic liquid-scintillator elements, covering the forward
1$\pi$ section of the solid angle around the target position. The
light charged particles (mainly protons and alphas) were detected by
the DIAMANT detector system consisting of 80 CsI scintillators
\cite{scheurer97,gal04}. Details of the experiment have been
described earlier \cite{cederwall11a}. Events were collected when at
least one neutron was detected by the Neutron Wall and one
$\gamma$-ray registered in coincidence in the clover detectors. With
these trigger conditions a total of 4 $\times$ 10$^9$ events were
recorded.

In the off-line processing, coincidence data were sorted into
symmetric $\gamma$-$\gamma$ matrices with different
conditions on the number of detected neutrons and charged particles. These conditions
were used to assign new $\gamma$-rays to $^{91}$Ru.
Coincidence $\gamma$-ray spectra were then obtained by setting gates in these
matrices.

\begin{figure*}[htbp]
\vspace{-4cm} \centering
\includegraphics[width=26cm,height=22cm,clip]{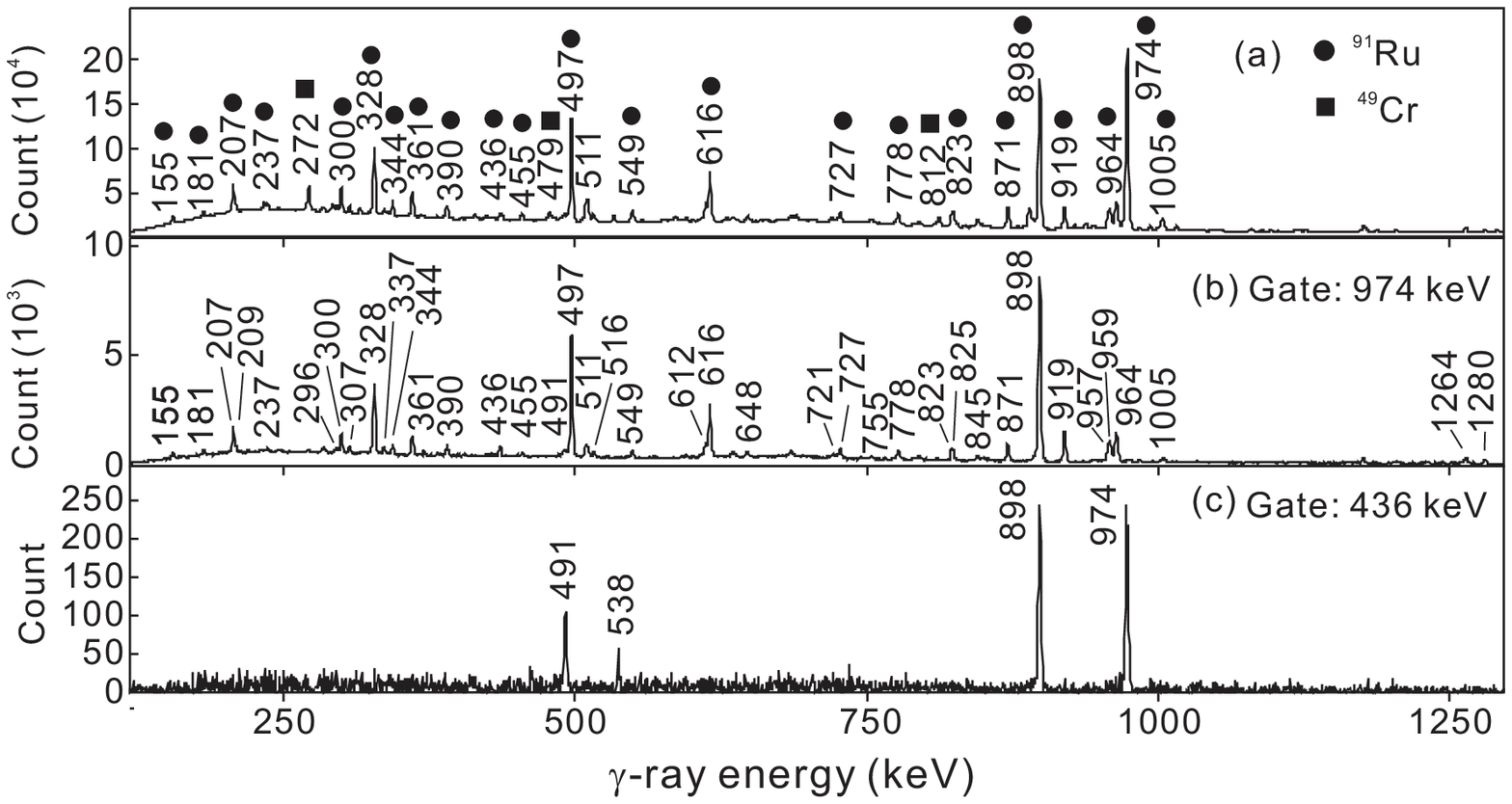}
\vspace{-8cm} \caption{(a) The total projection of a
$\gamma$-$\gamma$ matrix obtained in coincidence with two protons
and one neutron; the $\gamma$-ray peaks are transitions in $^{91}$Ru
and $^{49}$Cr. (b) A background subtracted spectrum of $\gamma$-rays
in coincidence with the 974 keV $\gamma$-ray, corresponding to the
transition that depopulates the (13/2$^{+}$) state in $^{91}$Ru. (c)
A spectrum gated on the 436 keV transition of $^{91}$Ru observed by
the present work.} \label{fig01}
\end{figure*}

Examples of coincidence spectra are shown in Fig. \ref{fig01}. Fig.
\ref{fig01}(a) is the total projection of a $\gamma$-$\gamma$ matrix
obtained in coincidence with the detection of two protons and one
neutron. This spectrum is dominated by $\gamma$-rays from $^{91}$Ru
with some peaks belonging to $^{49}$Cr produced in the
$^{16}$O($^{36}$Ar,$2p1n$) reaction, i.e. in the same reaction
channel. This contamination is removed by setting an additional
selection of known $\gamma$-rays in $^{91}$Ru. Fig. \ref{fig01}(b)
shows the spectrum obtained after gating on the 974 keV transition
previously known as deexciting the first excited state to the ground
state in $^{91}$Ru. This spectrum contains only known transitions
belonging to $^{91}$Ru with some additional, unknown $\gamma$-rays.
Further gating on these new transitions has allowed us to confirm
their assignment to $^{91}$Ru and to position them into the level
scheme. This is what is shown in Fig. \ref{fig01}(c) which gives a
spectrum gated on the new 436 keV transition of $^{91}$Ru. Finally,
with the large statistics obtained during this experiment it is
possible to perform a more detailed analysis of the observed
transitions. The geometry of the EXOGAM array allowed the
assignments of spins from the information on DCO ratios
\cite{krane73}. For this purpose, an asymmetric particle-gated
matrix was constructed in which $\gamma$ events recorded at
90$^{\circ}$ were sorted against those recorded at 135$^{\circ}$.
The experimental DCO ratios ($R_{\rm DCO}$) were deduced from pairs
of gated spectra according to equation
\begin{equation}
R_{\rm DCO}=\frac{I(\gamma_{1} \ \rm at \ 135^{\circ}; \ \rm gated \
by \ \gamma_{2} \ \rm at \ 90^{\circ})}{I(\gamma_{1} \ \rm at \
90^{\circ}; \ \rm gated \ by \ \gamma_{2} \ \rm at \
135^{\circ})}.\label{eq1}
\end{equation}

\begin{figure*}
\vspace{-5cm} \centering
\includegraphics[width=18cm,height=24cm,clip]{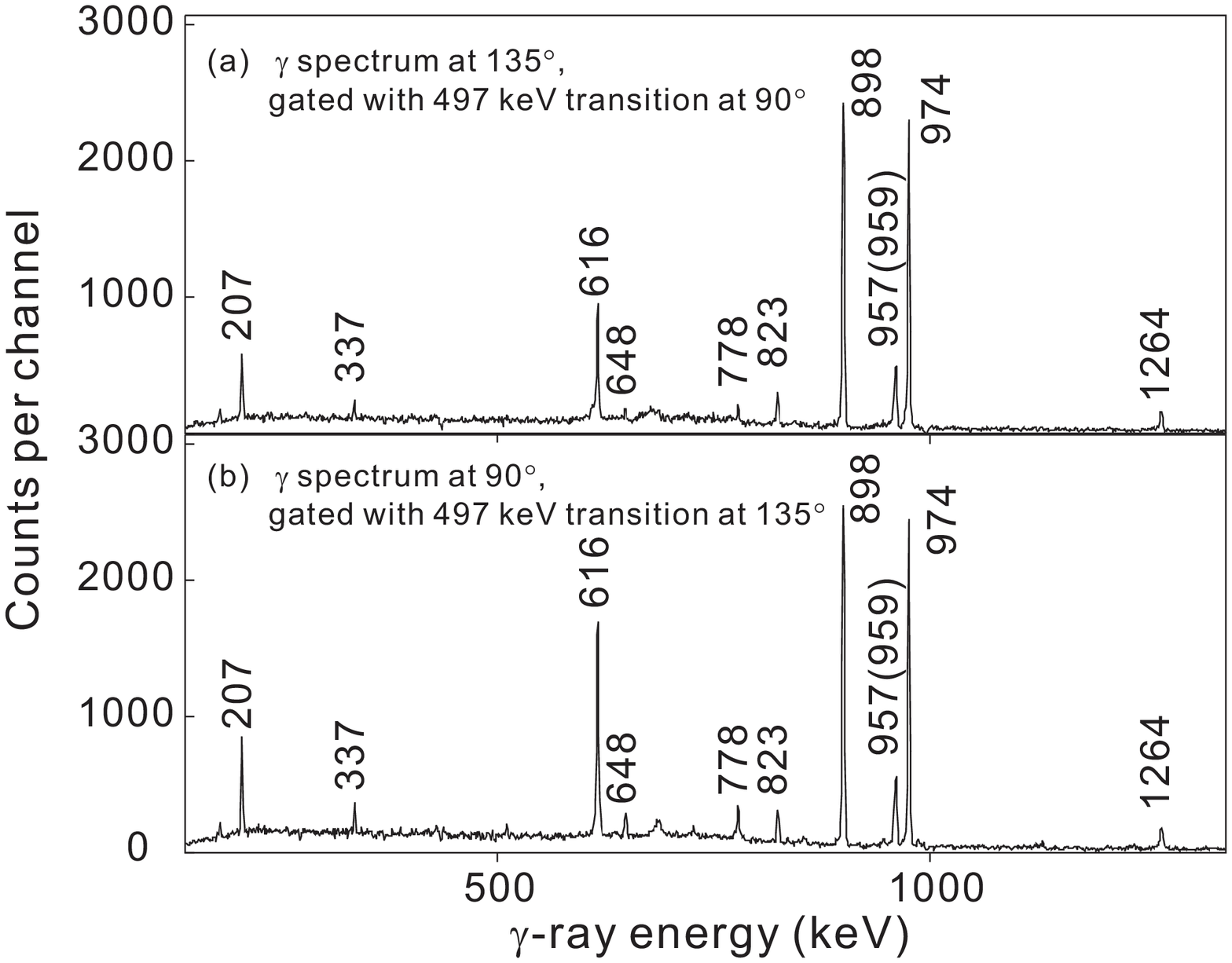}
\vspace{-7cm} \caption{(a) The projection of the DCO matrix on the
135$^\circ$ axis in coincidence with the 497 keV transition of
$^{91}$Ru at 90$^\circ$; and (b) the projection on the 90$^\circ$
axis in coincidence with the same transition at 135$^\circ$.}
\label{fig02}
\end{figure*}

\begin{figure*}[htbp]
\vspace{-0.4cm} \centering
\includegraphics[width=17cm,height=12cm,clip]{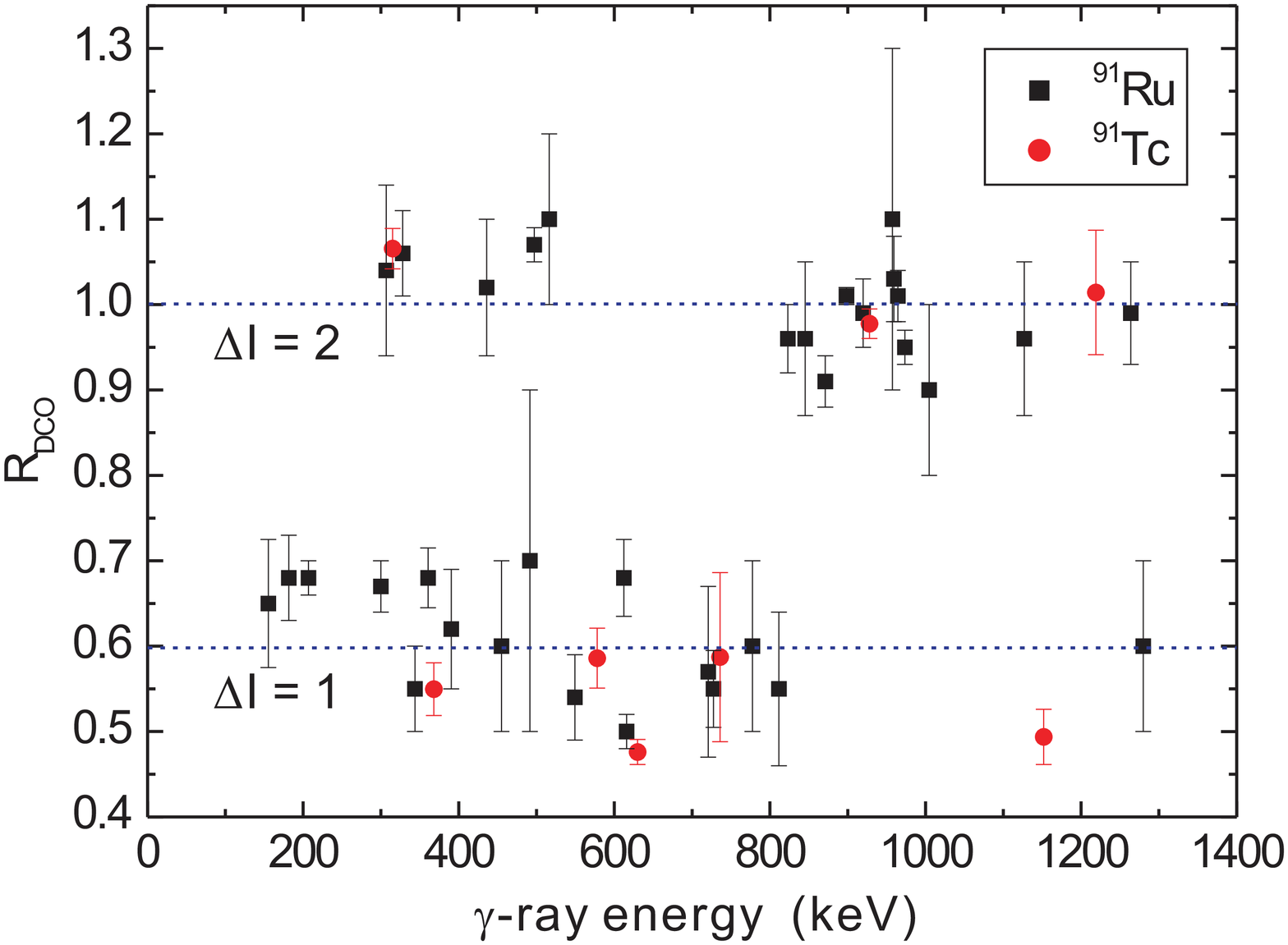}
\vspace{-0.5cm} \caption{(Color online) Experimental DCO ratios for
the transitions belonging to $^{91}$Ru ($\blacksquare$) and
$^{91}$Tc ({{\color{red}\huge$\bullet$}}). The lines correspond to
the values obtained for known dipole and quadrupole transitions
using gates on stretched quadrupoles and have been drawn to guide
the eye.} \label{fig03}
\end{figure*}

The detection efficiencies of detectors at 90$^{\circ}$ and at
135$^{\circ}$ have the same behaviour with $\gamma$-ray energy.
Therefore their ratio $R_{\rm eff}$ is a constant ($R_{\rm eff} =
1.79 \pm 0.05$) hence no efficiency correction of the DCO ratios was
needed. Fig. \ref{fig02} shows two projected spectra obtained from
the DCO matrix. The spectrum in the upper (lower) panel corresponds
to $\gamma$-rays detected at 135$^{\circ}$ (90$^{\circ}$) and in
coincidence with the 497 keV $(21/2^{+})\rightarrow(17/2^{+})$
transition in $^{91}$Ru observed at 90$^{\circ}$ (135$^{\circ}$).
The ratio of the peak intensities in these two spectra provides the
$R_{\rm DCO}$ values of the $\gamma$-rays. For example, the three
most intense transitions in both projected spectra shown in Fig.
\ref{fig02} are the 616, 898 and 974 keV $\gamma$-rays. Their
intensities in the two spectra are [2414(62), 6879(90), 6731(88)]
and [4821(80), 7134(92), 7062(90)], respectively. The deduced DCO
ratios for these transitions are then [0.50(2), 0.96(2), 0.95(2)].
The DCO ratios measured for $\gamma$-rays in $^{91}$Ru and also in
$^{91}$Tc produced in the $3p$ channel are shown in Fig.
\ref{fig03}. The $R_{\rm DCO}$ value for known stretched quadrupole
transitions is $\sim1$ and is $\sim0.6$ for known pure stretched
dipoles, when gating on quadrupole transitions. If the gate is set
on a pure stretched dipole transition, then the $R_{\rm DCO}$ value
for known quadrupole transitions is $\sim1.6$ and is $\sim1$ for
known pure stretched dipoles. Based on these assignment criteria,
the $R_{\rm DCO}$ values obtained in the above example suggest that
the 616 keV transition is a ${\Delta}I=1$ dipole transition whereas
the 898 and 974 keV transitions have a ${\Delta}I=2$ quadrupole
character. These assignments are consistent with the previous
assignments \cite{arnell93,heese94}. It should be noted that for
mixed $M1+E2$ transitions $R_{\rm DCO}$ ratios can vary between 0.6
and 1.0 depending on the $\delta$ multipole mixing ratio of the
$\gamma$-ray. A further ambiguity arises for non-stretched
(${\Delta}I=0$) pure $E1$ (or $M1$) transitions, where $R_{\rm DCO}$
for non-stretched dipole transition with $\delta\approx0$ mixing
ratio is approximately the same as for a stretched quadrupole
transition \cite{piip96,sohler12}. These ambiguities can be resolved
by simultaneously measuring the linear polarization of the
$\gamma$-ray transitions (see below). For example, stretched $E1$,
$E2$ or unstretched $M1$ transitions and stretched $M1$ or
unstretched $E1$ transitions have opposite sign linear polarization
values \cite{sohler12}.

In order to determine the multipolarity and the electromagnetic
nature of a transition, both the DCO ratio and the linear
polarization should be measured. One of the unique capabilities of
the EXOGAM array is the possibility to use the clover detectors as
Compton polarimeters. In the following, the measurement of the
performances of EXOGAM as Compton polarimeter will be described. The
clover detectors placed at 90$^\circ$ relative to the beam axis were
used to determine the linear polarization of $\gamma$-ray
transitions, since they are the most sensitive to the polarization
\cite{schlitt94}. We define the emission plane by the direction of
the initial $\gamma$-ray and the beam axis. The clover detector is
composed of four HPGe crystals closely packed in the same cryostat.
In Compton scattering, the initial and scattered $\gamma$-rays can
be detected in adjacent crystals of the same detector and analyzed
separately according to whether the scattering has occurred
horizontally to the emission plane or vertically to it.

\begin{figure*}
\vspace{-4cm} \centering
\includegraphics[width=17cm,height=23cm,clip]{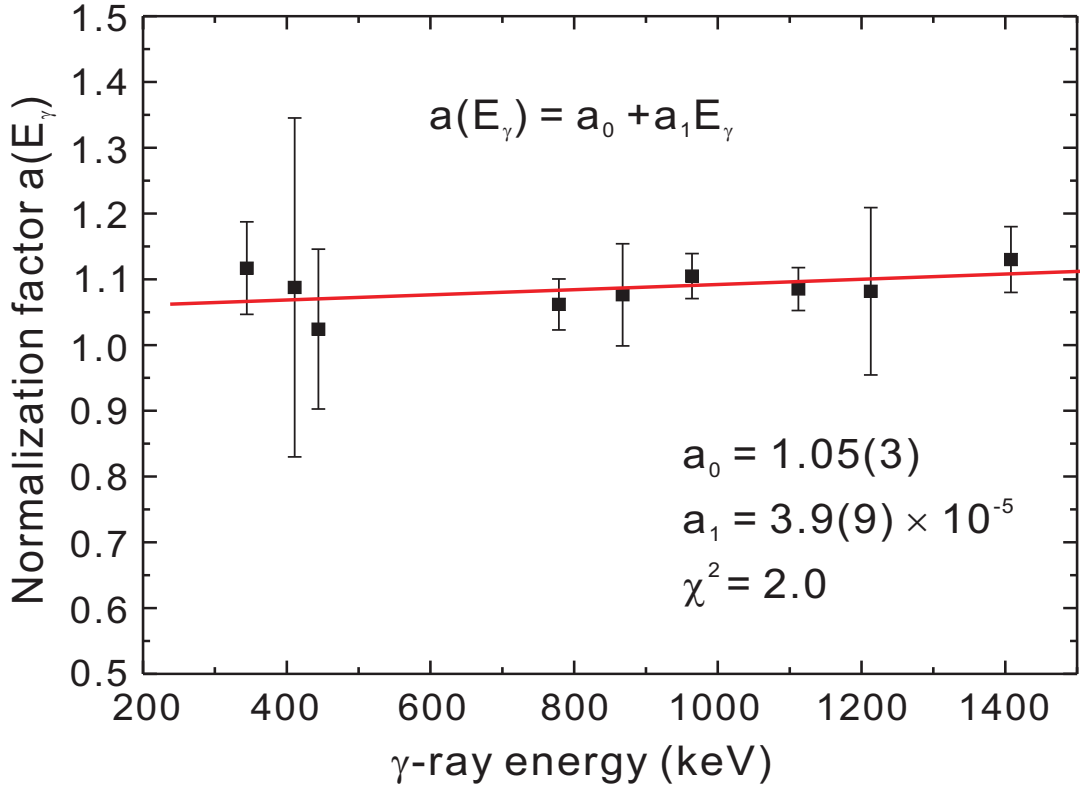}
\vspace{-8cm} \caption{(Color online) Normalization factor $a$ in
the linear polarization measurements as a function of $\gamma$-ray
energy $\left(\rm E_{\gamma}\right)$ for the EXOGAM
array.}\label{fig04}
\end{figure*}

Two $\gamma$-$\gamma$ matrices were created as follows: the first
$\gamma$-ray corresponds to a single-crystal hit in any clover
detector of the array and the second one to the sum of the energy
deposited in two crystals within the same clover located at
90$^\circ$ (i.e. the addback energy of events scattering between two
adjacent crystals of a clover, this one being positioned at
90$^\circ$). The matrices contain therefore events with either
horizontally or vertically scattered $\gamma$-rays in a clover at
90$^\circ$ on one axis and a single-crystal hit on any of the clover
detectors on the other axis. The number of horizontal ($N_{\bot}$)
and vertical ($N_{\|}$) scatters for a given $\gamma$-ray could be
obtained by setting gates on $\gamma$-ray transitions in the two
asymmetric matrices. The experimental polarization asymmetry is
defined by the ratio $A$,
\begin{equation}
A = \frac{\left[a\left(\rm
E_{\gamma}\right)N_{\bot}\right]-N_{\|}}{\left[a\left(\rm
E_{\gamma}\right)N_{\bot}\right]+N_{\|}},\label{eq2}
\end{equation}
where $a$(E$_{\gamma}$) is the normalization factor corresponding to
the asymmetry of the EXOGAM clover detectors, and is defined as
\begin{equation}
a\left(\rm E_{\gamma}\right) = \frac{N_{\|}\left(\rm
unpolarized\right)}{N_{\bot}\left(\rm
unpolarized\right)}.\label{eq3}
\end{equation}
The normalization factor is a function of $\gamma$-ray energy and
has been obtained from the measurement with a standard $^{152}$Eu
radioactive source. Fig. \ref{fig04} shows the variation of $a$ with energy
E$_{\gamma}$. It was fitted with the expression $a\left(\rm
E_{\gamma}\right)=a_{0}+a_{1}\rm E_{\gamma}$, resulting in
$a_{0}=1.05(3)$ and $a_{1}=3.9(9){\times}10^{-5}$, where
E$_{\gamma}$ is in keV. As is clear from Fig. \ref{fig04}, the value
of $a$ is almost constant and close to unity, showing nearly ideal
symmetry of the four-crystal clover detector acting as the Compton
polarimeter.

The polarization asymmetry $A$ is negative for unmixed stretched
magnetic transitions and positive for stretched electric
transitions. It is proportional to the degree of linear polarization
$P$,
\begin{equation}
A = QP,\label{eq4}
\end{equation}
where the quality factor $Q$ is the polarization sensitivity of the
polarimeter. $Q=0$ and $Q=1$ would indicate completely insensitive
and completely sensitive polarimeters, respectively. For a
point-like polarimeter, the polarization sensitivity $Q$ can be
calculated from the Klein-Nishina formula \cite{klein29}, which
gives
\begin{equation}
Q_{\rm point} = \frac{1+\alpha}{1+\alpha+\alpha^{2}} \ \ \rm with \
\ \alpha=\frac{\rm E_{\gamma}}{m_{e}c^{2}},\label{eq5}
\end{equation}
where m$_{e}$ is the electron rest mass. For a realistic setup of
detectors with finite crystal size, we have to integrate over a
certain range of scattering angles leading to a considerable
reduction of the polarization sensitivity. The effective
polarization sensitivity is usually given as
\begin{equation}
Q = Q_{\rm point} \cdot (p_{0}+p_{1}\rm E_{\gamma}) \label{eq6}
\end{equation}
According to eqs.(\ref{eq4})-(\ref{eq6}), $Q$ and the two parameters
$p_{0}$ and $p_{1}$ can be experimentally determined using
$\gamma$-rays whose linear polarization is well known. Theoretical
values of the linear polarization can be deduced from the angular
distribution. For $\gamma$-rays detected at 90$^{\circ}$ with
respect to the beam direction, the polarization of pure electric
quadrupole transitions can be calculated from the formula
\begin{equation}
P(90^{\circ}) = \frac{12A_{2}+5A_{4}}{8-4A_{2}+3A_{4}},\label{eq7}
\end{equation}
where $A_{2}$ and $A_{4}$ are the normalized ($A_{0}=1$)
coefficients of the Legendre polynomials in the angular
distribution.

\begin{figure*}
\vspace{-3.5cm} \centering
\includegraphics[width=16cm,height=22cm]{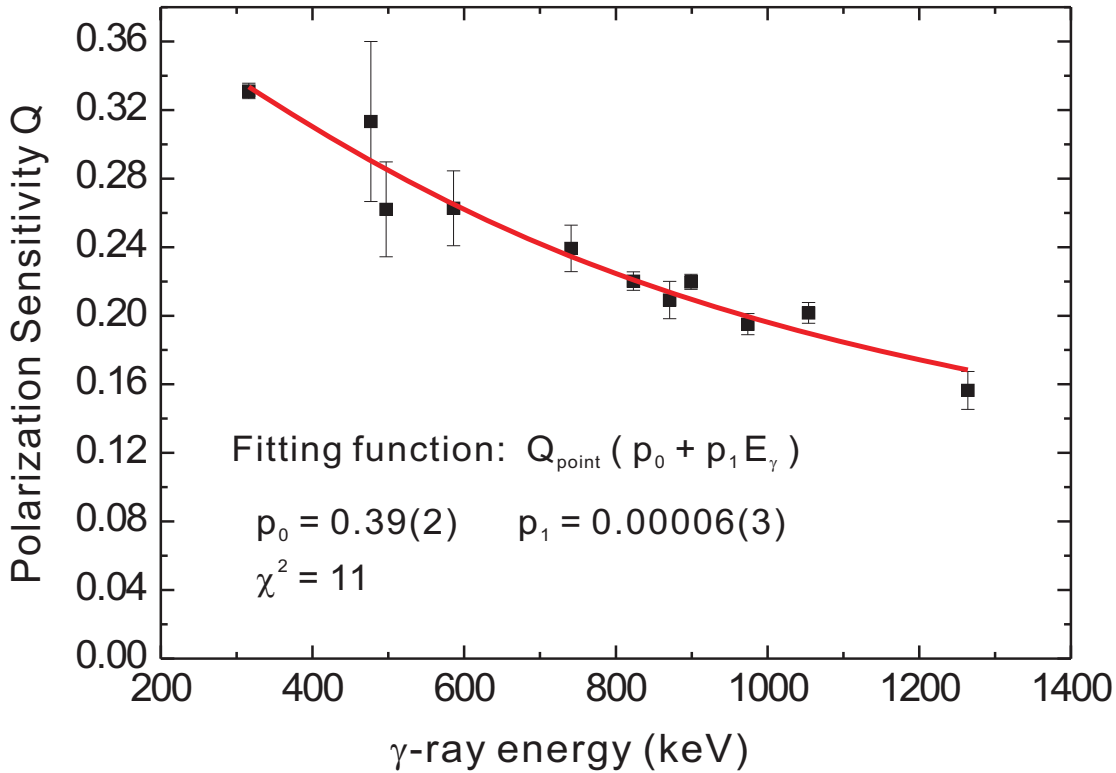}
\vspace{-8cm}
 \caption{(Color online) Polarization sensitivity of the EXOGAM
detector. The solid line is the fit to our data.} \label{fig05}
\end{figure*}

To determine the capability of the EXOGAM array to measure linear
polarization, we analyzed the angular distribution for the known
pure $E2$ transitions in the energy range 316 to 1264 keV in the
level schemes of $^{91}$Ru \cite{heese94}, $^{91}$Tc
\cite{rudolph94}, $^{90}$Mo \cite{singh92}, and $^{88}$Mo
\cite{weiszflog92}. The angular distribution coefficients, $A_{2}$
and $A_{4}$, for each transition, were extracted from least squares
fits of the photopeak areas and summarized in Table \ref{tab01}. The
deduced values of the linear polarization $P$ and the experimental
asymmetry ratio $A$ for the known $\gamma$-ray transitions are also
summarized in Table \ref{tab01}, along with the polarization
sensitivity $Q$ of the EXOGAM array. The experimental values of $Q$,
together with the results of the fit to the data, are shown in Fig.
\ref{fig05}. The coefficients $p_{0}$ and $p_{1}$ were determined by
a least squares fit to the values of $Q$ using the function of
eq.(\ref{eq6}); $p_{0}=0.39\pm0.02$, $p_{1}=0.00006\pm0.00003$.

The quality of a polarimeter depends on both its sensitivity to the
polarization and the coincidence efficiency between the scatterer
and absorber crystals expressed as \cite{Simpson83}
\begin{equation}
\epsilon_{\rm c}(E_{\gamma}) = \frac{N_{\bot}+ N_{\|}}{2N_{Clo}}
\cdot \epsilon_{Clo}(E_{\gamma}), \label{eq8}
\end{equation}
where $N_{Clo}$ and $\epsilon_{Clo}$ are the total number of counts
and the photopeak efficiency of the clover considered as a single
detector at the energy $E_{\gamma}$, measured when the $\gamma$-ray
has no polarization (i.e. using a source or when the detector is at
$0^{\circ}$ with respect to the beam direction). Finally it is
common to compare polarimeters using a figure of merit defined as
\cite{Simpson83}
\begin{equation}
{F} = Q^2 \cdot \epsilon_{\rm c} \label{eq9}
\end{equation}
The figure of merit deduced for the EXOGAM clover at 1368 keV is
1.51 $\times 10^{-5}$ which is 4.4 times larger that the one
measured for the smaller EUROGAM clover \cite{Clover}. At the same
$\gamma$-ray energy, the measured polarization sensitivity $\rm
Q_{EXOGAM}$ is 0.135(5) (0.121(5) for the EUROGAM clover) which
means that the increase is due to the much larger coincidence
efficiency. This increase in efficiency makes EXOGAM an ideal
polarimeter for low-intensity $\gamma$-rays.

\begin{table*}[H]
\begin{ruledtabular}
\caption{$\gamma$-ray energies, measured asymmetries, normalized
angular distribution coefficients, deduced $\gamma$-ray
polarization, and calculated polarization sensitivity of the EXOGAM
clover Ge detectors. (Only known pure $E2$ transitions have been
used to determine the polarization sensitivity $Q$, see text.)}
\label{tab01}
\begin{tabular}{ccccccccc}
 $E_{\gamma}$ (keV) & Channel & Nucleus & $J^{\pi}_{i}$$\rightarrow$$J^{\pi}_{f}$ & Asymmetry & $A_{2}$ & $A_{4}$  &
$P$ & $Q$  \\[1mm]
\hline \\[0.001mm]
316 & $3p$ & $^{91}$Tc & $21/2^{+} \rightarrow 17/2^{+}$ & 0.17(2) &
0.336(2) &
-0.178(2) & 0.513(2)  & 0.331(8)   \\
477 & $4p$ & $^{90}$Mo & $12^{+} \rightarrow 10^{+}$ & 0.16(4) &
0.32(8) &
-0.10(1) & 0.51(9)  & 0.31(4)   \\
497 & $2p1n$ & $^{91}$Ru & $21/2^{+} \rightarrow 17/2^{+}$ & 0.17(3)
& 0.39(7) &
-0.16(8) & 0.65(8)  & 0.26(3)   \\
586 & $2p1\alpha$ & $^{88}$Mo & $8^{+} \rightarrow 6^{+}$ & 0.11(3)
& 0.27(2) &
-0.08(3) & 0.42(4)  & 0.26(2)   \\
741 & $2p1\alpha$ & $^{88}$Mo & $2^{+} \rightarrow 0^{+}$ & 0.068(7)
& 0.22(2) &
-0.15(3) & 0.28(5)  & 0.24(2)   \\
823 & $2p1n$ & $^{91}$Ru & $25/2^{+} \rightarrow 21/2^{+}$ &
0.073(6) & 0.205(5) &
-0.02(1) & 0.33(1)  & 0.220(6)   \\
871 & $2p1n$ & $^{91}$Ru & $25/2^{-} \rightarrow 21/2^{-}$ &
0.081(9) & 0.23(3) &
-0.01(1) & 0.39(6)  & 0.21(1)   \\
898 & $2p1n$ & $^{91}$Ru & $17/2^{+} \rightarrow 13/2^{+}$ &
0.131(6) & 0.33(1) &
-0.01(2) & 0.60(3)  & 0.220(4)   \\
974 & $2p1n$ & $^{91}$Ru & $13/2^{+} \rightarrow 9/2^{+}$ & 0.139(4)
& 0.39(3) &
-0.02(5) & 0.71(8)  & 0.195(6)   \\
1054 & $4p$ & $^{90}$Mo & $4^{+} \rightarrow 2^{+}$ & 0.114(5) &
0.32(3) &
-0.02(1) & 0.57(6)  & 0.201(6)   \\
1264 & $2p1n$ & $^{91}$Ru & $25/2^{+} \rightarrow 21/2^{+}$ &
0.12(2) & 0.41(4) &
-0.01(1) & 0.8(1)  & 0.16(1)   \\
\end{tabular}
\end{ruledtabular}
\end{table*}

\section{Results}

The $\gamma$-rays from $^{91}$Ru were selected using the condition
that two protons and one neutron were detected and with an
additional selection on the two most intense transitions in
$^{91}$Ru. $\gamma$-ray energies, intensity, DCO and asymmetry
ratios have been measured (see Table \ref{tab02}). Spins and
parities of the levels have been assigned on the basis of the DCO
ratios and the linear polarization results, respectively.

\begin{figure*}[htbp]
\vspace{-1cm}
\centering
\includegraphics[width=16cm,height=12cm,,clip]{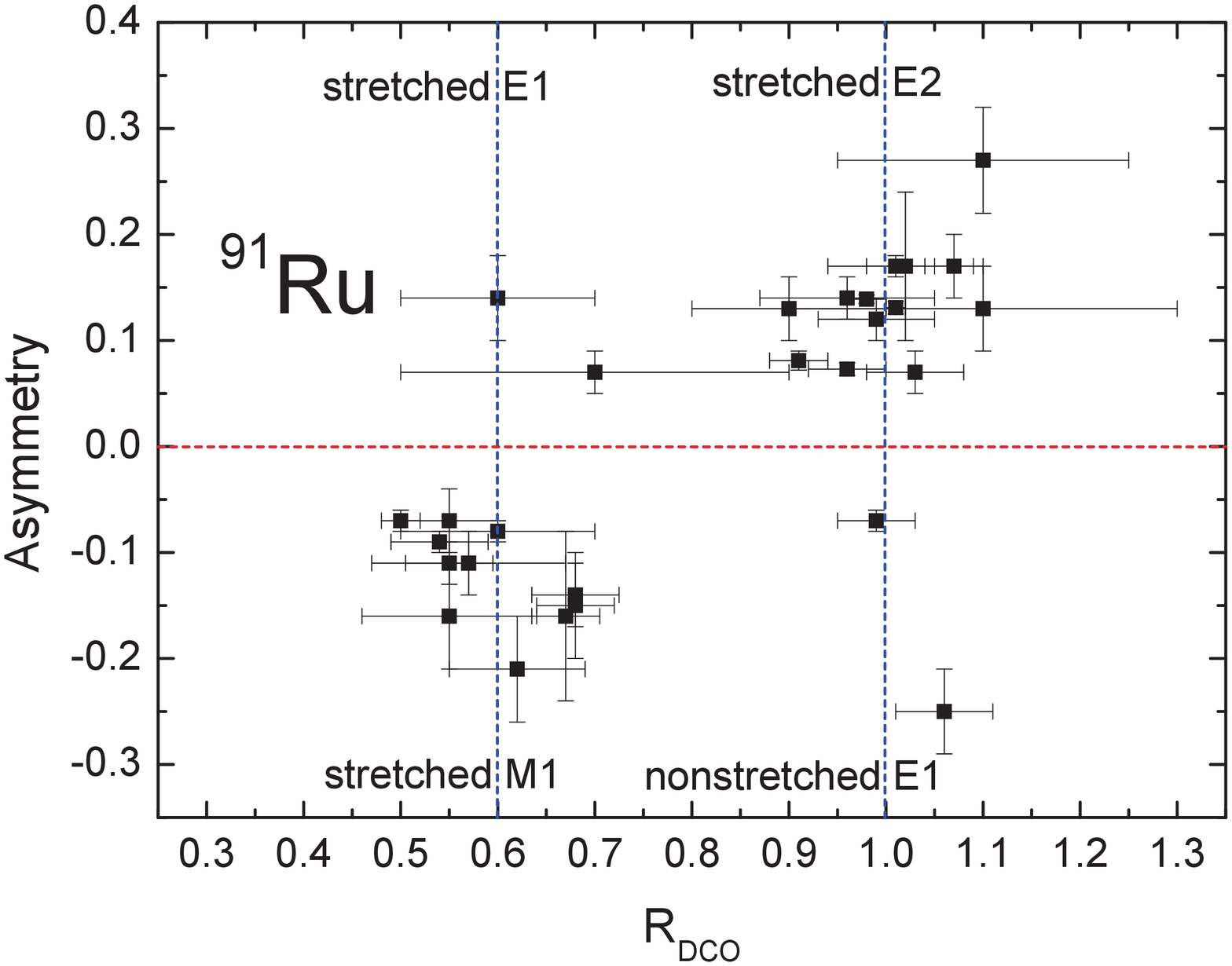}
\vspace{-0.5cm} \caption{(Color online) Two dimensional plot of the
asymmetry ratio $A$ as a function of the DCO ratio ($R_{\rm DCO}$)
of the $\gamma$-rays belonging to $^{91}$Ru. Stretched $E1$, $E2$
and $M1$ transitions and non-stretched $E1$ transitions are
indicated in the plot. The dashed lines parallel to the $y$-axis
correspond to the value obtained for known pure stretched dipole and
quadrupole transitions. These lines have been drawn to guide the
eye. The $R_{\rm DCO}$ values have been obtained after gating on a
quadrupole transition.} \label{fig06}
\end{figure*}

Fig. \ref{fig06} illustrates a two dimensional plot of the asymmetry
parameter $A$ as a function of the DCO ratio when gating on a
quadrupole transition. As can be seen from the plot, the
polarization and multipolarity measurements together give us a
reasonable assignment of the spin and parity for the levels.

\begin{figure*}[htbp]
\vspace{-6cm}
 \centering
\includegraphics[width=18cm,height=24cm]{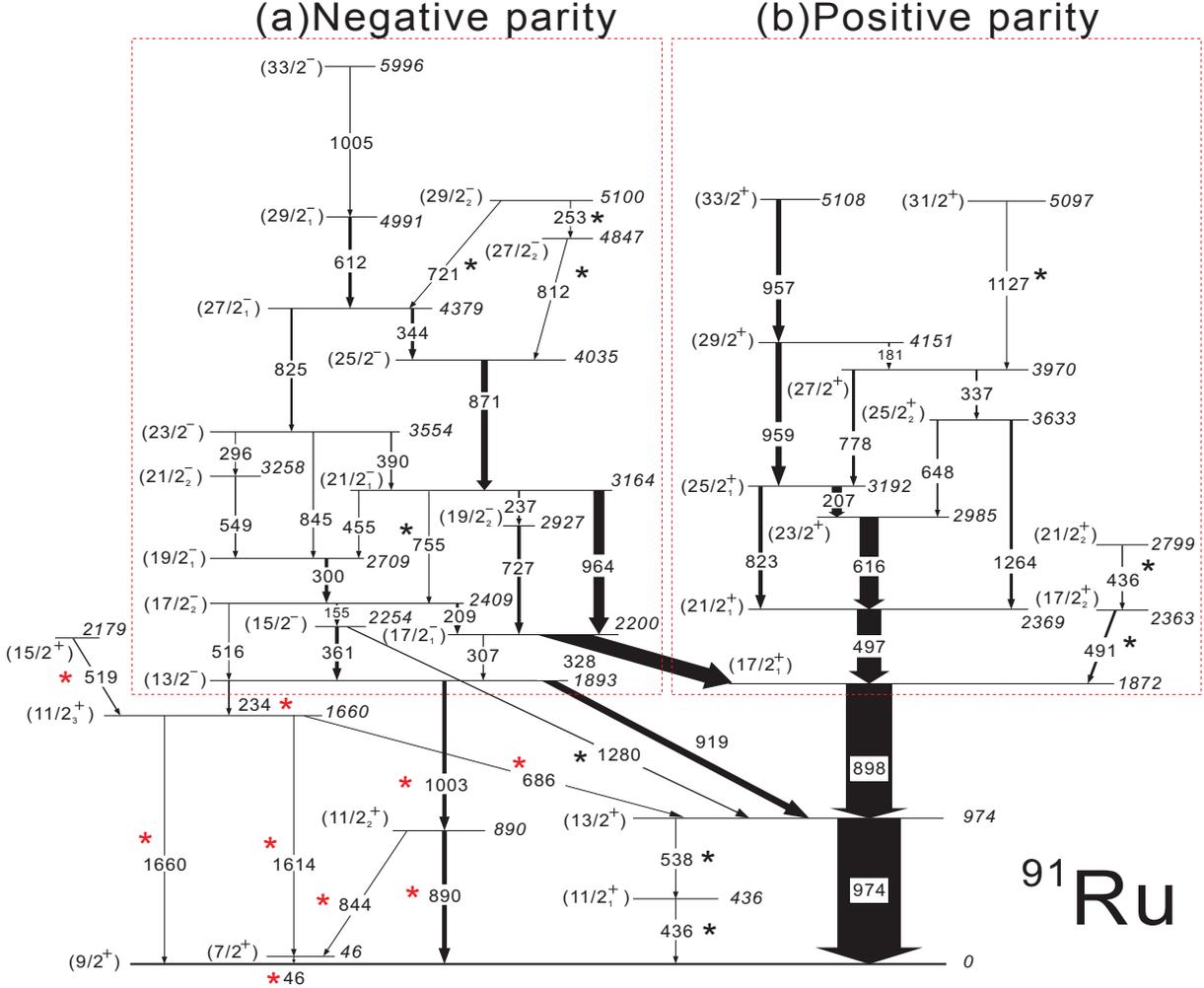}
\vspace{-5cm}
 \caption{(Color online) Level scheme of $^{91}$Ru proposed in the
present work. The new transitions are indicated by asterisks. Those
indicated with a red one will be discussed in a forthcoming paper.}
\label{fig07}
\end{figure*}

The deduced level scheme of $^{91}$Ru is shown in Fig. \ref{fig07}.
States above spin (33/2) seen in \cite{heese94} using the same
reaction channel could not be observed in our data because of the
lower beam energy (111 MeV compared to 149 MeV). The analysis of our
data revealed several new states. The ordering of the transitions in
the level scheme are fixed either with the help of some crossover
transitions or from the consideration of intensity balances in the
gated spectra. The analysis of the low-level structure below the
(13/2$^{-}$) state will be discussed in a forthcoming paper. In the
present one, we focus on transitions indicated with a black asterisk
in the level scheme of Fig. \ref{fig07}.

In $^{91}$Ru the ground state has been assumed to be (9/2$^{+}$)
\cite{arnell93,heese94, dean04}. This assumption is well supported
by the decay study of $^{91}$Ru \cite{komninos83a,hagberg83} and the
systematics of odd-A, $N=47$ isotones with 36 $\leqslant$ $Z$
$\leqslant$ 42 \cite{lederer78}. Above the ground state, a strong
transition sequence consisting of 974 keV, 898 keV, 497 keV, 823
keV, 959 keV, and 957 keV $\gamma$-rays was observed. The DCO ratio
analysis indicates that they are quadrupole transitions. The
polarization asymmetries for these quadrupole transitions are
clearly positive, showing that they are stretched $E2$ transitions
and thus have been assigned as de-exciting the positive-parity
levels as shown in the level scheme. A weak cascade of $\gamma$-rays
with energies of 538 keV and 436 keV has been assigned to the
present level scheme as parallel to the 974 keV,
(13/2$^{+}$)$\rightarrow$(9/2$^{+}$) transition. The ordering of
these two transitions is based on their relative intensities. In
addition, the DCO ratio analysis and polarization measurement show
that the weak 538 keV transition has a $M1$ character, leading to
the assignment of ($11/2^{+}$) for the new yrast level at 436 keV.
In the $\beta$-decay of $^{91}$Rh \cite{gorska00} several
transitions were observed and assigned to feed the ground-state of
$^{91}$Ru. Their placement in the level scheme is not confirmed in
\cite{dean04} but our measurement confirms the excited states at 436
keV and 890 keV. It is also noted that the 436 keV line is a doublet
(see later).

Above the excited state at 974 keV, the level scheme is separated
into two parts. One part is the group of positive-parity states
which is on the right-hand side of the level scheme (Fig.
\ref{fig07}(b)). The other part is the group of states on the
left-hand side of the level scheme (Fig. \ref{fig07}(a)). Since no
linear polarization measurement has been performed for this latter
group of states up to now, the negative parity assignment proposed
for those states in the earlier works \cite{arnell93,heese94} could
only be based on indirect evidence and hence was only tentative. Of
primary importance in the linear polarization measurements are the
most intense $\gamma$-ray transitions connecting the low-lying
positive-parity states (17/2$^{+}_{1}$) and (13/2$^{+}$) of the
yrast band and the presumed negative-parity levels. In $^{91}$Ru the
key transitions for determining the parity of the left-side
structure are the 328 keV, 919 keV and 1280 keV lines. From the
results of the DCO ratio ($\sim1$ when gated by the stretched
quadrupole transitions) and linear polarization measurements
($A<0$), a non-stretched $\Delta$$I=0$ $E1$ character (i.e. parity
change) for the 328 keV and 919 keV connecting transitions is
obtained. Thus these two $\gamma$-rays have been assigned as the
(13/2$^{-}$)$\rightarrow$(13/2$^{+}$) and
(17/2$^{-}_{1}$)$\rightarrow$(17/2$^{+}_{1}$) transitions,
respectively. From the $M1$ and $E2$ character and multipolarity of
the transitions depopulating levels above the (13/2$^{-}$) and
(17/2$^{-}_{1}$) states lying at 1893 keV and 2200 keV respectively,
negative parity has been assigned to these states. The DCO ratio and
asymmetry measured for the 1280 keV transition are respectively
0.6(1) and 0.14(4) indicating an $E1$ character, which is consistent
with the previous assignments.

Up to the 5996 keV state, our spin assignments of the
negative-parity level structure confirm the proposed values of
Refs.\cite{arnell93,heese94} but, from intensity considerations, the
ordering of the 296 keV and 549 keV transitions is changed. The 549
keV transition in the sequence depopulates the 3258 keV state and
feeds the $J^{\pi}=(19/2^{-}_{1}$), 2709 keV state. This transition
is a stretched magnetic dipole, and thus allows the assignment of
$J^{\pi}=(21/2^{-}_{2}$) to the state at 3258 keV. The observation
of the new $\gamma$-rays of 812 keV, 253 keV, and 721 keV lying
above the (25/2$^{-}$) state at 4035 keV establishes two states as
shown in Fig. \ref{fig08}. These two states, which are connected by
the 253 keV transition, de-excite via the 812 keV and 721 keV
$\gamma$-rays to the $J^{\pi}=(25/2^{-}$), 4035 keV and
$J^{\pi}=(27/2^{-}_{1}$), 4379 keV states, respectively. The
combination of the DCO ratio and linear polarization data determines
the multipolarities of the 812 keV and 721 keV $\gamma$-rays to be
both stretched $M1$. Therefore, spin and parity of (27/2$^{-}_{2}$)
are assigned for the 4847 keV level and (29/2$^{-}_{2}$) for the
5100 keV level. This is further supported by the stretched dipole
character of the 253 keV transition obtained from the DCO ratio
analysis.

For the assignments of the positive-parity states, up to the
(33/2$^{+}$) level at 5108 keV, our results are consistent with the
earlier work of Refs.\cite{arnell93,heese94}. In addition, three new
transitions of 436 keV, 491 keV and 1127 keV have been observed
below the (33/2$^{+}$) state. From the spectrum gated on the 436 keV
peak shown in Fig. \ref{fig01}(c), the 436 keV transition is only in
coincidence with the 491 keV line and the most intense 974 keV-898
keV transition sequence. The 538 keV line shown in this spectrum is
in coincidence with the other 436 keV doublet transition and has
been placed in the level scheme as feeding the new (11/2$^{+}_{1}$)
state. Therefore, the 491 keV-436 keV cascade is proposed to be
built directly on the (17/2$^{+}_{1}$) state at 1872 keV. The
ordering of these two new transitions is determined from the
relative intensities in the coincidence spectra. Based on the
results of the DCO ratio and linear polarization measurements, the
491 and 436 keV $\gamma$-rays have been assigned as
(17/2$^{+}_{2}$)$\rightarrow$(17/2$^{+}_{1}$) and
(21/2$^{+}_{2}$)$\rightarrow$(17/2$^{+}_{2}$) transitions,
respectively. This results in the determination of the second
(17/2$^{+}_{2}$) and (21/2$^{+}_{2}$) states at 2363 and 2799 keV,
respectively. The weak 1127 keV transition populating the
$J^{\pi}=(27/2^{+}$), 3970 keV state shows a possible $E2$
character; thus, the spin and parity of (31/2$^{+}$) is tentatively
assigned to the 5097 keV state depopulated by the 1127 keV
transition. We stress that except for this latter (31/2$^{+}$)
state, it is {\it only} the ground state spin and parity uncertainty
that needs resolving to allow all our assignments to be firmly
established.

\section{Semi-empirical shell-model calculation and discussion}

\begin{figure*}[htbp]
\vspace{-2cm}
 \centering
\includegraphics[width=12cm,height=17cm]{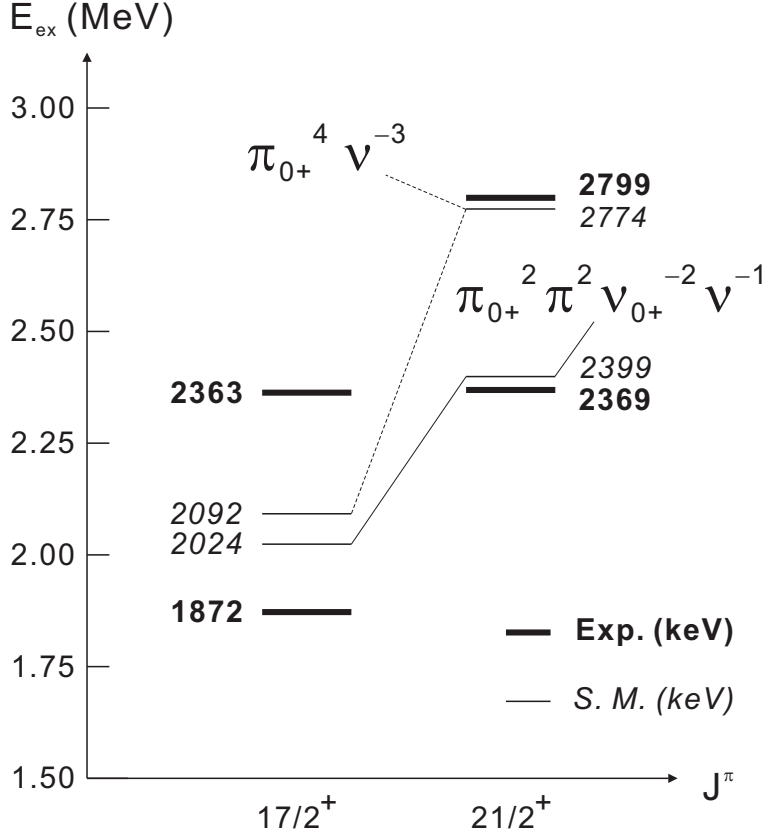}
\vspace{-3cm} \caption{Comparison between the calculated (thin
lines) and experimentally observed (solid bars) (17/2$^{+}$) and
(21/2$^{+}$) states in $^{91}$Ru. See text for the configuration
notation.} \label{fig08}
\end{figure*}

To better understand the microscopic structure of the states of
interest, a semi-empirical shell model was used. This allows the
calculation of the excitation energy of complex multi-particle-hole
configurations from the excitation energies of known configurations
in neighboring nuclei. This method is parameter independent and was
proposed by Garvey and Kelson \cite{garvey66,garvey69} for
ground-state masses based on the prescription by Talmi and de Shalit
\cite{talmi62,shalit63}. The technique was later extended by
Blomqvist and collaborators \cite{blomqvist83} to calculate excited
states in the $A$ $\sim$ 150 and 200 mass regions. The approach
restricts the analysis to states that predominantly contain a pure
single-particle configuration as is expected for most of the yrast
or near-yrast levels. We will mainly discuss the yrast and
near-yrast seniority-three states. These level energies are
calculated using nuclear ground-state masses, single-particle
energies and two-particle interactions obtained from experimental
data. The calculated results are compared with the experimental
observations in Fig. \ref{fig08} for the first and second
(17/2$^{+}$) and (21/2$^{+}$) states. Input data for the
calculations are taken from the neighbouring nuclei $^{85,87,88}$Sr
\cite{arnell77,ekstrom81} $^{87-90}$Zr \cite{kitching78,bendahan86},
$^{90-92}$Mo \cite{kabadiyski92,ray04,singh92}, and $^{93,94}$Ru
\cite{komninos83b,mills07}. Ground-state masses needed in the
calculations are obtained from Ref. \cite{mass03}.

The non yrast states with $J^{\pi}=(17/2^{+}_{2}$) and (21/2$^{+}_{2}$) have
been identified in $^{91}$Ru and added to the level scheme. As
already mentioned previously, the simplest low-lying excitations
expected for $^{91}$Ru are those arising from the
$\nu$$g_{9/2}^{-3}$ configuration terminating at spin 21/2$^{+}$.
However, as $g_{9/2}$ protons are active, a different seniority
scheme involving proton excitations, such as two $g_{9/2}$ protons
coupled to the unpaired $g_{9/2}$ neutron hole, might become yrast
in this nucleus. This $\pi$$g_{9/2}^{2}$$\nu$$g_{9/2}^{-1}$
multiplet terminates at spin 25/2$^{+}$. In $^{91}$Mo, the
three-quasiparticle seniority-three
($\pi$$g_{9/2}^{2}$$\nu$$g_{9/2}^{-1}$)21/2$^{+}$ and 17/2$^{+}$
configurations were assigned to the 2268 and 2069 keV states, respectively
\cite{ray04}.
Since the only active nucleons are $g_{9/2}$ proton(s) and/or neutron(s), we will simplify the
notation and omit the explicit reference to the $g_{9/2}$ single-particle level. We will only
specify the pairs that are coupled to $0^{+}$ and the total angular momentum $J_{Tot}$ when
applicable i.e.:
$(\pi ^{i}_{0^+} \pi ^{j} \nu ^{k}_{0^+} \nu ^{l})J_{Tot}$ means {\it i} protons in $g_{9/2}$ coupled to
$0^{+}$, {\it j} protons in $g_{9/2}$ {\it not} coupled to $0^{+}$, the same for the {\it k} and {\it l}
neutrons, the total angular momentum being $J_{Tot}$. For instance the
$(\pi (g_{9/2})^{2}_{0^+}(g_{9/2})^{2} \nu (g_{9/2})^{-2}_{0^+}(g_{9/2})^{-1})21/2^{+}$
configuration will be reduced to ($\pi$$_{0^{+}}^{2}$$\pi$$^{2}$$\nu$$_{0^{+}}^{-2}$$\nu$$^{-1}$)21/2$^{+}$.

The energies of the seven-quasiparticle seniority-three $(\pi
^{2}_{0^+} \pi ^{2} \nu ^{-2}_{0^+} \nu ^{-1})$21/2$^{+}$ and
17/2$^{+}$ levels in $^{91}$Ru can be calculated from the
above-mentioned $\pi^{2}\nu^{-1}$ states in $^{91}$Mo (see
\cite{blomqvist83, piip96} for the details). For example, with the
known excitation energies of the concerned configurations in
neighboring nuclei, the energy of the $(\pi ^{2}_{0^+} \pi ^{2} \nu
^{-2}_{0^+} \nu ^{-1})$21/2$^{+}$ state is calculated as

\begin{eqnarray*}
E_{(\pi ^{2}_{0^+} \pi ^{2} \nu ^{-2}_{0^+} \nu ^{-1})21/2^{+}}^{ \ \rm ^{91}Ru} & =
& E_{(\pi ^{2} \nu ^{-1}) 21/2^{+}}^{ \ \rm ^{91}Mo}\\
& + & \frac{23}{30}(E_{(\pi ^{2}_{0^+} \pi ^{2})8^{+}}^{ \
\rm ^{94}Ru}+E_{(\pi ^{2} \nu ^{-2}_{0^+})8^{+}}^{ \
\rm ^{90}Mo}-2E_{(\pi ^{2})8^{+}}^{ \ \rm ^{92}Mo}) \\
& + & \frac{7}{30}(E_{(\pi ^{2}_{0^+} \pi ^{2})6^{+}}^{ \
\rm ^{94}Ru}+E_{(\pi ^{2} \nu ^{-2}_{0^+})6^{+}}^{ \
\rm ^{90}Mo}-2E_{(\pi ^{2})6^{+}}^{ \ \rm ^{92}Mo})+S \\
& = & 2399 \  \rm keV.
\end{eqnarray*}
The fractions in the formula are angular momentum recoupling
coefficients; in this reduction the mass term $S$ is
\begin{eqnarray*}
S & = & 2M_{\rm ^{91}Mo}+2M_{\rm ^{90}Mo}-4M_{\rm ^{92}Mo}+M_{\rm
^{87}Zr}+3M_{\rm ^{90}Zr}-2M_{\rm ^{88}Zr}-2M_{\rm ^{89}Zr}+M_{\rm
^{94}Ru}-M_{\rm ^{91}Ru} \\
& = & -99 \ \rm keV.
\end{eqnarray*}
The energy of the 17/2$^{+}$ state with the same configuration is
calculated in a similar way to be 2024 keV. The calculated energies
of the seven-quasiparticle seniority-three $(\pi ^{2}_{0^+} \pi ^{2}
\nu ^{-2}_{0^+} \nu ^{-1})$21/2$^{+}$ and 17/2$^{+}$ states are
comparable to those of the yrast (21/2$^{+}_{1}$) and
(17/2$^{+}_{1}$) levels observed at 2369 and 1872 keV, respectively,
so we suggest that these levels have the dominant configuration of
$(\pi ^{2}_{0^+} \pi ^{2} \nu ^{-2}_{0^+} \nu ^{-1})$. This is
consistent with the calculated results for these two levels in the
previous work \cite{heese94}.

In $^{85}$Sr \cite{arnell77}, the five-quasiparticle
seniority-three $(\pi ^{-2}_{0^+} \nu ^{-3})$21/2$^{+}$ and 17/2$^{+}$
configurations were identified at 3082 and 2400 keV. Therefore, the
seven-quasiparticle seniority-three
$(\pi ^{4}_{0^+} \nu ^{-3})$ 21/2$^{+}$ and 17/2$^{+}$
states might be expected in $^{91}$Ru. The excitation energy of the
$(\pi ^{4}_{0^+} \nu ^{-3})$ 21/2$^{+}$ state is
calculated to be
\begin{eqnarray*}
E_{(\pi ^{4}_{0^+} \nu ^{-3})}^{ \ \rm ^{91}Ru}
& = & E_{(\pi ^{-2}_{0^+} \nu ^{-3}) 21/2^{+}}^{ \ \rm ^{85}Sr}+
3E_{(\pi ^{4}_{0^+} \nu ^{-1}) 9/2^{+}}^{ \ \rm ^{93}Ru}-
3E_{(\pi ^{-2}_{0^+} \nu ^{-1}) 9/2^{+}}^{ \ \rm ^{87}Sr}-
2E_{(\pi ^{4}_{0^+})}^{ \ \rm ^{94}Ru}+S \\
& = & 2774 \ \rm keV.
\end{eqnarray*}
In this case, the mass term $S$ is
\begin{eqnarray*}
S & = & M_{\rm ^{85}Sr}-M_{\rm ^{91}Ru}+3M_{\rm ^{93}Ru}-3M_{\rm
^{87}Sr}+2M_{\rm ^{88}Sr}-2M_{\rm ^{94}Ru}=-308 \ \rm keV.
\end{eqnarray*}
A similar calculation gives an energy of 2092 keV for the
$(\pi ^{4}_{0^+} \nu ^{-3})$17/2$^{+}$ configuration.
The (21/2$^{+}_{2}$) and (17/2$^{+}_{2}$) levels are observed at 2799
and 2363 keV, and their energies are close to the calculated values
of the $(\pi ^{4}_{0^+} \nu ^{-3})$21/2$^{+}$ and
17/2$^{+}$ configurations. Therefore, the experimentally observed
(21/2$^{+}_{2}$) and (17/2$^{+}_{2}$) states might be associated with
the $J_{\rm max}$ and $J_{\rm max}-2$ members of the
seven-quasiparticle seniority-three
$(\pi ^{4}_{0^+} \nu ^{-3})$ multiplet.
It is noted that these two states decay to the
(17/2${_1}^{+}$) state via the weak 491 keV $\gamma$-ray.

\section{Summary}

In the present work, we have used the EXOGAM Ge clover detectors as
Compton polarimeters to measure the linear polarization of
$\gamma$-ray transitions observed in $^{91}$Ru. The polarization
sensitivity of the EXOGAM clover detectors has been obtained for
incident $\gamma$-ray energies ranging from 300 to 1300 keV. Using
the DCO ratios and linear polarization measurements, the nature and
multipolarity of the transitions of interest have been deduced.
However, since the ground state spin and parity in $^{91}$Ru is not
yet measured, only the tentative spins and parities have been
assigned to the yrast and non yrast states in $^{91}$Ru. We stress
that resolving the ground state spin and parity would allow the firm
assignment of all the identified levels except the (31/2$^{+}$)
state. New (21/2$^{+}_{2}$) and (17/2$^{+}_{2}$) states have been
observed at 430 keV and 491 keV above the yrast (21/2$^{+}_{1}$) and
(17/2$^{+}_{1}$) states, respectively. Semi-empirical shell-model
calculations have been done for these yrast and non yrast levels.
The results clearly reveal the characteristic features of the active
protons and neutrons in the $g_{9/2}$ orbital. The $(\pi ^{2}_{0^+}
\pi ^{2} \nu ^{-2}_{0^+} \nu ^{-1})$21/2$^{+}$ and 17/2$^{+}$
configurations are proposed for the yrast (21/2$^{+}$) and
(17/2$^{+}$) levels, and the $(\pi ^{4}_{0^+} \nu ^{-3})$21/2$^{+}$
and 17/2$^{+}$ configurations are assigned to the non-yrast
(21/2$^{+}_{2}$) and (17/2$^{+}_{2}$) levels.

\section{ACKNOWLEDGMENT}

The authors would like to thank the operators of the GANIL
cyclotrons for providing the $^{36}$Ar beam. We would also like to
thank the EXOGAM collaboration for use of the clover Ge detector
array, the DIAMANT collaboration for use of the charged particle
detector system, and the European $\gamma$-ray Spectroscopy Pool for
use of the neutron detector system. We acknowledge funding support
from the French-Polish LEA COPIGAL and the
IN2P3-Polish laboratories COPIN agreement number 06-122,
from the UK STFC,
from the Swedish Research Council (contract nos 2007-4067 and 2008-5793), from
the G\"{o}ran Gustafsson Foundation, from the OTKA under contract number K100835 and from
the Bolyai J\'anos Foundation.~A.~G. has been supported by the  Generalitat Valenciana,
Spain, under grant PROMETEO/2010/101 and by MINECO, Spain, under grants AIC-D-2011-0746
and FPA2011-29854.~A.~J acknowledge financial support
from the Spanish Ministerio de Ciencia e Innovaci\'{o}n under contract FPA2011-29854-C04.
~Z.~Y. acknowledges the support from the Chinese Academy of Sciences, China.



\newpage

\renewcommand{\baselinestretch}{1.0}

\begin{table*}[tbp]
\begin{ruledtabular}
\caption{Properties of the $\gamma$-rays of $^{91}$Ru, produced in
the $^{58}$Ni($^{36}$Ar,{\it 2p1n})$^{91}$Ru reaction. Uncertainties are
given in parentheses. The gates used for the determination of the DCO
ratios are indicated in the table.} \label{tab02}
\begin{tabular}{ccccccc}
 $E_{\gamma}$ (keV) \footnote[1]{Energy uncertainties are within 0.5 keV.} & $I_{\gamma}$ ($\%$) \footnote[2]{$\gamma$-ray intensities relative to the
$(13/2^{+})\rightarrow(9/2^{+})$ 974 keV transition.} & $E_{i}$
$\rightarrow$ $E_{f}$ & $J^{\pi}_{i}$ $\rightarrow$ $J^{\pi}_{f}$ &
$R_{\rm DCO}$  &
Gate$_{\rm DCO}$ (keV) & Asymmetry  \\[1mm]
\hline \\[0.001mm]
 155.4 & 3.0(2) & 2409 $\rightarrow$ 2254 & (17/2$_{2}$$^{-}$) $\rightarrow$ (15/2$^{-}$) & 0.65(7) & 974 &  \\
 181.6 & 2.7(1) & 4151 $\rightarrow$ 3970 & (29/2$^{+}$) $\rightarrow$ (27/2$^{+}$) & 0.68(5) & 974 &  \\
 206.9 & 16.0(4) & 3192 $\rightarrow$ 2985 & (25/2$_{1}$$^{+}$) $\rightarrow$ (23/2$^{+}$) & 0.68(2) & 497 &  \\
 209.4 & 4.5(2) & 2409 $\rightarrow$ 2200 & (17/2$_{2}$$^{-}$) $\rightarrow$ (17/2$_{1}$$^{-}$) &  &  &  \\
 236.8 & 2.9(3) & 3164 $\rightarrow$ 2927 & (21/2$_{1}$$^{-}$) $\rightarrow$ (19/2$_{2}$$^{-}$) &  &  &  \\
 252.9 & $<$0.6 & 5100 $\rightarrow$ 4847 & (29/2$_{2}$$^{-}$) $\rightarrow$ (27/2$_{2}$$^{-}$) &  &  &  \\
 296.0 & 1.6(2) & 3554 $\rightarrow$ 3258 & (23/2$^{-}$) $\rightarrow$ (21/2$_{2}$$^{-}$) &  &  &  \\
 299.9 & 5.6(2) & 2709 $\rightarrow$ 2409 & (19/2$_{1}$$^{-}$) $\rightarrow$ (17/2$_{2}$$^{-}$) & 0.67(3) & 974 & -0.16(8) \\
 306.8 & 1.9(2) & 2200 $\rightarrow$ 1893 & (17/2$_{1}$$^{-}$) $\rightarrow$ (13/2$^{-}$) & 1.04(10) & 974 &  \\
 328.0 & 25.1(1) & 2200 $\rightarrow$ 1872 & (17/2$_{1}$$^{-}$) $\rightarrow$ (17/2$_{1}$$^{+}$) & 1.06(5) & 898 & -0.25(4) \\
 336.5 & 3.2(2) & 3970 $\rightarrow$ 3633 & (27/2$^{+}$) $\rightarrow$ (25/2$_{2}$$^{+}$) &  &  &  \\
 343.8 & 5.2(1) & 4379 $\rightarrow$ 4035 & (27/2$^{-}$) $\rightarrow$ (25/2$^{-}$) & 0.55(5) & 974 & -0.07(3) \\
 360.6 & 5.9(2) & 2254 $\rightarrow$ 1893 & (15/2$^{-}$) $\rightarrow$ (13/2$^{-}$) & 0.68(4) & 974 & -0.15(5) \\
 390.5 & 2.7(2) & 3554 $\rightarrow$ 3164 & (23/2$^{-}$) $\rightarrow$ (21/2$_{1}$$^{-}$) & 0.62(7) & 974 & -0.21(5) \\
 435.9 & 2.4(2) & 2799 $\rightarrow$ 2363 & (21/2$_{2}$$^{+}$) $\rightarrow$ (17/2$_{2}$$^{+}$) & 1.02(8) & 974 & 0.17(7) \\
 436.0 & $<$0.4 & 436 $\rightarrow$ 0 & (11/2$_{1}$$^{+}$) $\rightarrow$ (9/2$^{+}$) &  &  &  \\
 455.0 & 1.0(1) & 3164 $\rightarrow$ 2709 & (21/2$_{1}$$^{-}$) $\rightarrow$ (19/2$_{1}$$^{-}$) & 0.6(1) & 974 &  \\
 491.4 & 4.2(2) & 2363 $\rightarrow$ 1872 & (17/2$_{2}$$^{+}$) $\rightarrow$ (17/2$_{1}$$^{+}$) & 0.7(2) & 974 & 0.07(2) \\
 497.2 & 38.3(1) & 2369 $\rightarrow$ 1872 & (21/2$_{1}$$^{+}$) $\rightarrow$ (17/2$_{1}$$^{+}$) & 1.07(2) & 974 & 0.17(3) \\
 516.4 & 1.1(1) & 2409 $\rightarrow$ 1893 & (17/2$_{2}$$^{-}$) $\rightarrow$ (13/2$^{-}$) & 1.1(1) & 974 & 0.27(5) \\
 538.0 & $<$0.4 & 974 $\rightarrow$ 436 & (13/2$^{+}$) $\rightarrow$ (11/2$_{1}$$^{+}$) & 1.1(6) & 361 & -0.19(8) \\
 549.3 & 2.4(2) & 3258 $\rightarrow$ 2709 & (21/2$_{2}$$^{-}$) $\rightarrow$ (19/2$_{1}$$^{-}$) & 0.54(5) & 974 & -0.09(1) \\
 612.3 & 5.4(2) & 4991 $\rightarrow$ 4379 & (29/2$^{-}$) $\rightarrow$ (27/2$^{-}$) & 0.68(4) & 871 & -0.14(3) \\
 615.8 & 30.0(5) & 2985 $\rightarrow$ 2369 & (23/2$^{+}$) $\rightarrow$ (21/2$_{1}$$^{+}$) & 0.50(2) & 497 & -0.07(1) \\
 648.0 & 2.8(2) & 3633 $\rightarrow$ 2985 & (25/2$_{2}$$^{+}$) $\rightarrow$ (23/2$^{+}$) &  &  &  \\
 720.7 & 0.7(1) & 5100 $\rightarrow$ 4379 & (29/2$_{2}$$^{-}$) $\rightarrow$ (27/2$^{-}$) & 0.57(10) & 871 & -0.11(3) \\
 727.5 & 5.8(3) & 2927 $\rightarrow$ 2200 & (19/2$_{2}$$^{-}$) $\rightarrow$ (17/2$_{1}$$^{-}$) & 0.55(4) & 974 & -0.11(2) \\
 754.5 & 1.7(2) & 3164 $\rightarrow$ 2409 & (21/2$_{1}$$^{-}$) $\rightarrow$ (17/2$_{2}$$^{-}$) &  &  &  \\
 777.5 & 4.8(1) & 3970 $\rightarrow$ 3192 & (27/2$^{+}$) $\rightarrow$ (25/2$_{1}$$^{+}$) & 0.6(1) & 974 & -0.08(1) \\
 811.6 & 0.8(1) & 4847 $\rightarrow$ 4035 & (27/2$_{2}$$^{-}$) $\rightarrow$ (25/2$^{-}$) & 0.55(9) & 871 & -0.16(5) \\
 823.0 & 6.8(2) & 3192 $\rightarrow$ 2369 & (25/2$_{1}$$^{+}$) $\rightarrow$ (21/2$_{1}$$^{+}$) & 0.96(4) & 497 & 0.073(6) \\
 824.7 & 3.6(2) & 4379 $\rightarrow$ 3554 & (27/2$^{-}$) $\rightarrow$ (23/2$^{-}$) &  &  &  \\
 845.3 & 2.0(1) & 3554 $\rightarrow$ 2709 & (23/2$^{-}$) $\rightarrow$ (19/2$_{1}$$^{-}$) & 0.96(9) & 974 & 0.14(2) \\
 871.2 & 10.5(1) & 4035 $\rightarrow$ 3164 & (25/2$^{-}$) $\rightarrow$ (21/2$_{1}$$^{-}$) & 0.91(3) & 974 & 0.081(9) \\
 898.5 & 73(1) & 1872 $\rightarrow$ 974 & (17/2$_{1}$$^{+}$) $\rightarrow$ (13/2$^{+}$) & 1.01(1) & 974 & 0.131(6) \\
 919.8 & 11.3(1) & 1893 $\rightarrow$ 974 & (13/2$^{-}$) $\rightarrow$ (13/2$^{+}$) & 0.99(4) & 974 & -0.07(1) \\
 957.4 & 8.1(3) & 5108 $\rightarrow$ 4151 & (33/2$^{+}$) $\rightarrow$ (29/2$^{+}$) & 1.1(2) & 1264 & 0.13(4) \\
 959.4 & 9.6(3) & 4151 $\rightarrow$ 3192 & (29/2$^{+}$) $\rightarrow$ (25/2$_{1}$$^{+}$) & 1.03(5) & 957 & 0.07(2) \\
 964.5 & 17.3(3) & 3164 $\rightarrow$ 2200 & (21/2$_{1}$$^{-}$) $\rightarrow$ (17/2$_{1}$$^{-}$) & 1.01(3) & 871 & 0.17(1) \\
 973.5 & 100 & 974 $\rightarrow$ 0 & (13/2$^{+}$) $\rightarrow$ (9/2$^{+}$) & 0.95(2) & 497 & 0.139(4) \\
 1004.7 & 1.7(1) & 5996 $\rightarrow$ 4991 & (33/2$^{-}$) $\rightarrow$ (29/2$^{-}$) & 0.9(1) & 974 & 0.13(3) \\
 1126.9 & 0.8(1) & 5097 $\rightarrow$ 3970 & (31/2$^{+}$) $\rightarrow$ (27/2$^{+}$) & 0.96(9) & 497 &  \\
 1263.9 & 4.9(2) & 3633 $\rightarrow$ 2369 & (25/2$_{2}$$^{+}$) $\rightarrow$ (21/2$_{1}$$^{+}$) & 0.99(6) & 974 & 0.12(2) \\
 1280.7 & 2.1(8) & 2254 $\rightarrow$ 974 & (15/2$^{-}$) $\rightarrow$ (13/2$^{+}$) & 0.6(1) & 974 & 0.14(4) \\
\end{tabular}
\end{ruledtabular}
\end{table*}


\begin{thebibliography} {99}

\bibitem{matthias75} E. Matthias, E. Recknagel, O. Hashimoto, S.
Nagamiya, K. Nakai, T. Yamasaki, and Y. Yamasaki, Nucl. Phys. A{\bf
237}, 182 (1975).

\bibitem{hausser78} O. H\"{a}usser, T. Faestermann, I. S. Towner,
T. K. Alexander, H. R. Andrews, J. R. Beene, D. Horn, D. Ward, and
C. Broude, Hyperfine Interact. {\bf 4}, 196 (1978).

\bibitem{weiszflog93} M. Weiszflog, D. Rudolph, C. J. Gross, M. K.
Kabadiyski, K. P. Lieb, H. Grawe, J. Heese, K. -H. Maier, J. Eberth,
and S. Skoda, Z. Phys. A {\bf 344}, 395 (1993).

\bibitem{zwar85} D. Zwarts, Comp. Phys. Commun. {\bf 38}, 365 (1985).

\bibitem{weiszflog95} M. Weiszflog, A. Jungclaus, D. Kast, K. P.
Lieb, R. Schubart, H. Grawe, J. Heese, and K. -H. Maier, Z. Phys.
A {\bf 353}, 7 (1995).

\bibitem{arnell93} S. E. Arnell, D. Foltescu, H. A. Roth,
\"{O}. Skeppstedt, A. Nilsson, S. Mitarai, and J. Nyberg, Phys.
Scr. {\bf 47}, 355 (1993).

\bibitem{heese94} J. Heese, H. Grawe, K. H. Maier, R. Schubart,
F. Cristancho, C. J. Gross, A. Jungclaus, K. P. Lieb, D. Rudolph,
J. Eberth, and S. Skoda, Phys. Rev. C {\bf 49}, 1896 (1994).

\bibitem{dean04} S. Dean, M. G\'{o}rska, F. Aksouh, H. de Witte,
M. Facina, M. Huyse, O. Ivanov, K. Krouglov, Yu. Kudryavtsev,
I. Mukha, D. Smirnov, J.-C. Thomas, K. Van de Velc, J. Van de Walle,
P. Van Duppen, and J. Van Roosbroeck,
Eur. Phys. J. A {\bf 21}, 243 (2004).

\bibitem{gorska00} M. G\'{o}rska, S. Dean, V. Prasad N.V.S., A. Andreyev,
B. Bruyneel, S. Franchoo, M. Huyse, K. Krouglov, R. Raabe, K. Van de Vel,
P. Van Duppen, and J. Van Roosbroeck,
Proceedings of the International Workshop PINGST2000, Selected Topics on N=Z nuclei,
June 2000, Lund, Sweden, Eds. D. Rudolph and M. Hellstr\"{o}m, (Bloms i Lund AB, 2000).

\bibitem{Clover} G. Duch\^{e}ne, F.A. Beck, P.J. Twin, G. de France, D. Curien,
L. Han, C.W. Beausang, M.A. Bentley, P.J. Nolan, J. Simpson,
Nucl. Instr. Meth. A {\bf 432}, 90 (1999).

\bibitem{Andgren07} K. Andgren, E. Ganio\v{g}lu, B. Cederwall, R. Wyss, S. Bhattacharyya,
 J. R. Brown, G. de Angelis, G. de France, Zs. Dombr\`{a}di, J. G\`{a}l, B. Hadinia,
A. Johnson, F. Johnston-Theasby, A. Jungclaus, A. Khaplanov, J.
Kownacki, K. Lagergren, G. La Rana, J. Moln\`{a}r, R. Moro, B. S.
Nara Singh, J. Nyberg, M. Sandzelius, J.-N. Scheurer, G. Sletten, D.
Sohler, J. Tim\`{a}r, M. Trotta, J. J. Valiente-Dob\'{o}n, E.
Vardaci, R. Wadsworth, and S. Williams , Phys. Rev. C {\bf 76},
014307 (2007).

\bibitem{exogam} J.Simpson, F. Azaiez, G. de France, J. Fouan, J. Gerl, R. Julin,
W. Korten, P.J. Nolan, G Sletten, P.M. Walker and the EXOGAM collaboration,
Acta Physica Hungarica, New Series, Heavy Ion Physics {\bf 11}, 159 (2000).

\bibitem{skeppstedt99} \"{O}. Skeppstedt, H. A. Roth, L.
Lindstr\"{o}m, R. Wadsworth, I. Hibbert, N. Kelsall, D.
Jenkins, H. Grawe, M. G\'{o}rska, M. Moszy\'{n}ski, Z.
Sujkowski, D. Wolski, M. Kapusta, M. Hellstr\"{o}m, S.
Kalogeropoulos, D. Oner, A. Johnson, J. Cederk\"{a}ll, W.
Klamra, J. Nyberg, M. Weiszflog, J. Kay, R. Griffiths, J. Garces
Narro, C. Pearson, and J. Eberth, Nucl. Instr. Meth. A {\bf 421}, 531 (1999).

\bibitem{scheurer97} J. N. Scheurer, M. Aiche, M. M. Aleonard, G.
Barreau, F. Bourgine, D. Boivin, D. Cabaussel, J. F. Chemin, T. P.
Doan, J. P. Goudour, M. Harston, A. Brondi, G. La Rana, R. Moro, E.
Vardaci, and D. Curien, Nucl. Instr. Meth. A {\bf 385}, 501 (1997).

\bibitem{gal04} J. G\'{a}l, G. Hegyesi, J. Moln\'{a}r,
B. M. Nyak\'{o}, G. Kalinka, J. N. Scheurer, M. M.
Al\'{e}onard, J. F. Chemin, J. L. Pedroz, K. Juh\'{a}sz,
and V. F. E. Pucknell, Nucl. Instr. Meth. A {\bf 516}, 502 (2004).

\bibitem{cederwall11a}  B. Cederwall, F. Ghazi Moradi, T. B\"{a}ck, A. Johnson, J. Blomqvist, E. Cl\'{e}ment, G.de France, R. Wadsworth, K. Andgren, K. Lagergren, A. Dijon, G. Jaworski, R. Liotta, C. Qi, B.M. Nyak\'{o}, J. Nyberg, M. Palacz, H. Al-Azri, A. Algora, G. de Angelis, A. Atac, S. Bhattacharyya, T. Brock, J.R. Brown, P. Davies, A. Di Nitto, Zs. Dombradi, A. Gadea, J. G\'{a}l, B. Hadinia, F. Johnston-Theasby, P. Joshi, K. Juh\'{a}sz, R. Julin, A. Jungclaus, G. Kalinka, S.O. Kara, A. Khaplanov, J. Kownacki, G. La Rana, S.M. Lenzi, J. Moln\'{a}r, R. Moro, D.R. Napoli, B.S. Nara Singh, A. Persson, F. Recchia, M. Sandzelius, J.-N. Scheurer, G. Sletten, D. Sohler, P.-A. Soderstrom, M.J. Taylor, J. Timar, J.J. Valiente-Dobon, E. Vardaci, and S. Williams, Nature {\bf 469}, 68 (2011).

\bibitem{krane73} K. S. Krane, R. M. Steffen, and R. M. Wheeler,
Nucl. Data Tables {\bf 11}, 351 (1973).

\bibitem{piip96} M. Piiparinen, A. Ata\c{c}, J. Blomqvist, G. B. Hagemann, B. Herskind, R. Julin, S. Juutinen, A. Lampinen, J. Nyberg, G. Sletten, P. Tikkanen, S. T\"{o}rm\"{a}nen, A. Virtanen, R. Wyss, Nucl. Phys. A {\bf 605}, 191 (1996).

\bibitem{sohler12} D. Sohler, I. Kuti, J. Tim\'{a}r, P. Joshi, J. Moln\'{a}r, E. S. Paul, K. Starosta, R. Wadsworth, A. Algora, P. Bednarczyk, D. Curien, Zs. Dombr\'{a}di, G. Duchene, D. B. Fossan, J. G\'{a}l, A. Gizon, J. Gizon, D. G. Jenkins, K. Juh\'{a}sz, G. Kalinka, T. Koike, A. Krasznahorkay, B. M. Nyak\'{o}, P. M. Raddon, G. Rainovski, J. N. Scheurer, A. J. Simons, C. Vaman, A. R. Wilkinson, and L. Zolnai, Phys. Rev. C {\bf 85}, 044303 (2012).

\bibitem{schlitt94} B. Schlitt, U. Maier, H. Friedrichs, S. Albers,
I. Bauske, P. von Brentano, R. D. Heil, R.-D. Merzberg, U. Kneissl,
J. Margraf, H. M. Pitz, C. Wesselborg and A. Zilges, Nucl. Instr.
Meth. A {\bf 337}, 416 (1994).

\bibitem{klein29} O. Klein, Y. Nishina, Z. Phys. {\bf 52}, 853 (1929).

\bibitem{rudolph94} D. Rudolph, C. J. Gross, A. Harder, M. K.
Kabadiyski, K. P. Lieb, M. Weiszflog, J. Altmann, A. Dewald, J.
Eberth, T. Mylaeus, H. Grawe, J. Heese, and K.-H. Maier, Phys. Rev.
C {\bf 49}, 66 (1994).

\bibitem{singh92} P. Singh, R. G. Pillay, J. A. Sheikh, and H.
G. Devare, Phys. Rev. C {\bf 45}, 2161 (1992).

\bibitem{weiszflog92} M. Weiszflog, K. P. Lieb, F. Cristancho, C. J.
Gross, A. Jungclaus, D. Rudolph, H. Grawe, J. Heese, K. -H. Maier,
R. Schubart, J. Eberth, and S. Skoda, Z. Phys. A {\bf 342}, 257 (1992).

\bibitem{Simpson83} J.Simpson, P.A. Butler, L.P. Ekstrom,
Nucl. Instr. Meth. A {\bf 204}, 463 (1983).

\bibitem{komninos83a} P. Komninos, E. Nolte, and P. Blasi, Z. Phys. {\bf A314}, 135 (1983).

\bibitem{hagberg83} E. Hagberg, J. C. Hardy, H. Schmeing, E. T. H.
Clifford, and V. T. Koslowsky, Nucl. Phys. A {\bf 395}, 152 (1983).

\bibitem{lederer78} C. M. Lederer, V. S. Shirley, "Table of
Isotopes, 7th Edition" (Wiley, New York 1978).

\bibitem{garvey66} G. T. Garvey, T. Kelson, Phys. Rev. Lett. {\bf 16}, 197 (1966).

\bibitem{garvey69} G. T. Garvey, Annu. Rev. Nucl. Sci. {\bf 19}, 433 (1969).

\bibitem{talmi62} I. Talmi, Rev. Mod. Phys. {\bf 34}, 704 (1962).

\bibitem{shalit63} A. de Shalit, I. Talmi, Nuclear Shell Theory (New York: Academic 1963).

\bibitem{blomqvist83} J. Blomqvist, P. Kleinheinz, and P. J. Daly, Z. Phys. A {\bf 312}, 27 (1983).

\bibitem{arnell77} S. E. Arnell, S. Sj\"{o}berg, \"{O}.
Skeppstedt, and E. Wallander, Nucl. Phys. {\bf A280}, 72 (1977).

\bibitem{ekstrom81} L. P. Ekstr\"{o}m, G. D. Jones, F. Kearns,
T. P. Morrison, A. Nilsson, V. Paar, P. J. Twin, R. Wadsworth, E.
Wallander, and N. J. Ward, J. Phys. G: Nucl. Phys. {\bf 7}, 85 (1981).

\bibitem{kitching78} J. E. Kitching, P. A. Batay-Csorba, C. A.
Fields, R. A. Ristinen, and B. L. Smith, Nucl. Phys. A {\bf 302}, 159 (1978).

\bibitem{bendahan86} J. Bendahan, C. Broude, E. Dafni, G. Goldring,
M. Hass, E. Naim, and M. H. Rafailovich, Phys. Rev. C {\bf 33}, 1517 (1986).

\bibitem{kabadiyski92} M. K. Kabadiyski, F. Cristancho, C. J. Gross,
A. Jungclaus, K. P. Lieb, D. Rudolph, H. Grawe, J. Heese, K. -H.
Maier, J. Eberth, S. Skoda, W. -T. Chou, and E. K. Warburton, Z.
Phys. A {\bf 343}, 165 (1992).

\bibitem{ray04} S. Ray, N. S. Pattabiraman, R. Goswami, S. S.
Ghugre, A. K. Sinha, and U. Garg, Phys. Rev. C {\bf 69}, 054314 (2004).

\bibitem{komninos83b} P. Komninos, E. Nolte, Z. Phys. A {\bf 310}, 137 (1983).

\bibitem{mills07} W. J. Mills, J. J. Ressler, R. A. E. Austin, R. S.
Chakrawarthy, D. S. Cross, A. Heinz, E. A. McCutchan, and M. D.
Strange, Phys. Rev. C {\bf 75}, 047302 (2007).

\bibitem{mass03} The 2012 Update to the Atomic Mass Evaluation.

\end{thebibliography}
\end{document}